\documentclass[preprint,5p,twocolumn,numbers]{elsarticle}

\usepackage{url}
\usepackage{amsmath}
\usepackage{balance}
\usepackage{graphicx}

\usepackage[colorlinks=true,
linkcolor=blue,citecolor=blue,urlcolor=blue]{hyperref}

\newcommand{\Qa}{$\textmd{RQ}_1$}
\newcommand{\Qb}{$\textmd{RQ}_2$}
\newcommand{\Qc}{$\textmd{RQ}_3$}

\usepackage{xcolor}

\usepackage{fancyhdr} 
\fancypagestyle{firststyle}
{
   \fancyhf{}
   \fancyfoot[C]{\scriptsize{\textit{Information Systems}, published online in August 2021, pp.~1--11. This is the authors' copy. The publisher's copy is available online at \url{https://doi.org/10.1016/j.is.2021.101876}.}}
}

\begin{document}

\title{The GDPR Enforcement Fines at Glance}

\author[utu]{Jukka Ruohonen\corref{cor}}
\ead{juanruo@utu.fi}
\author[utu]{Kalle Hjerppe}
\cortext[cor]{Corresponding author.}
\address[utu]{Department of Future Technologies, University of Turku, FI-20014 Turun yliopisto, Finland}

\begin{abstract}
The General Data Protection Regulation (GDPR) came into force in 2018. After
this enforcement, many fines have already been imposed by national data
protection authorities in Europe. This paper examines the individual GDPR
articles referenced in the enforcement decisions, as well as predicts the amount
of enforcement fines with available meta-data and text mining features extracted
from the enforcement decision documents. According to the results,
three articles related to the general principles, lawfulness, and
information security have been the most frequently referenced ones. Although the
amount of fines imposed vary across the articles referenced, these three
particular articles do not stand out. Furthermore, a better statistical evidence
is available with other meta-data features, including information about the
particular European countries in which the enforcements were made. Accurate
predictions are attainable even with simple machine learning techniques for
regression analysis. Basic text mining features outperform the meta-data
features in this regard. In addition to these results, the paper reflects the
GDPR's enforcement against public administration obstacles in the European Union
(EU), as well as discusses the use of automatic decision-making systems in
judiciary.
\end{abstract}

\begin{keyword}
Data protection, privacy, law enforcement, public administration, legal mining,
empirical jurisprudence
\end{keyword}

\maketitle

\section{Introduction}

\thispagestyle{firststyle} 

Data protection has a long history in Europe~\cite{Flaherty86}.\footnote{~This
paper is an extended version of an earlier conference paper presented at COUrT
-- CAiSE for legal documents workshop~\cite{Ruohonen20COURT}.} With respect to
the EU, the GDPR repealed the earlier Directive 95/46/EC. Although this
directive laid down much of the legal groundwork for EU-wide data protection,
its national adaptations, legal interpretations, and enforcement varied both
across the member states and different EU institutions~\cite{Erdos16, Fuster14,
  Ventrella20}. In short: it was a paper tiger. Later on, the provisions for
both privacy and data protection were strengthened by the inclusion of them in
the Charter of Fundamental Rights of the European Union (EU), signed with the
Treaty of Lisbon in 2009. The GDPR is the latest manifestation in this
path: the goal of the regulation is to protect natural persons with respect to
the processing of their personal data, and, therefore, the goal is also to guard
their fundamental right to data protection.

The GDPR has been extensively studied in recent years. To put political,
economic, and related reasons aside, the reason for the abundance of research
originates from the regulation's scope. The fifth Article~(A) defines personal
data as any information relating to an identified or identifiable natural
person. Thus, with few restrictions, as specified in A23 and A89, the GDPR
covers all processing activities of personal data, whether manual or
automated. This wide scope means that it is difficult to consider the regulation
without a context. The protection of personal data is different for information
systems than it is for biomedical applications; it differs between scholarly
disciplines, from computer science to medicine. The GDPR establishes only a few
general principles that are universal. As specified in A5, these include
lawfulness, fairness, and transparency, purposefulness, data minimization,
accuracy, finite data retention, integrity, confidentiality, and
accountability. It is possible to derive design patterns from these
principles~\cite{Hjerppe19a, Shastri21}, but the patterns are still dependent on
a given context. By implication, it is impossible to establish universal
guidelines with which sanctions could be avoided. This provides a motivation for
the present work to examine the specific articles that have been referenced by
data protection authorities (DPAs) when imposing fines according to the
conditions specified in A83.

Another motivation stems from the noted administration and governance issues for
European data protection practices. Akin to some other public administration
domains, such as product safety administration~\cite{Ruohonen21SF}, the history
of the European data protection has always relied heavily on the ombudsmen-like
DPAs instead of enforcement through litigation or criminal law~\cite{Dalenius79,
  Hustinx09}. However, a reasonably comprehensive literature search indicates no
previous empirical research on the enforcement of this particular regulation,
excluding an earlier conference paper~\cite{Ruohonen20COURT} upon which the
present paper builds. Compared to the conference paper, the present work
presents a more thorough examination of the enforcement fines, including the
prediction of these by text mining techniques and regression analysis. The
predictions are also discussed with respect to a broader debate on automatic
decision-making (ADMs) systems used in the public sector. In addition, the work
extends the examination toward the GDPR's administrative and political
aspects. To these ends, the present paper examines the following three Research
Questions (RQs) regarding the enforcement fines:
\begin{description}
\itemsep 2pt
\item{\Qa:~\textit{Which GDPR articles have been actively referenced in
    the recent enforcement cases?}}
\item{\Qb:~\textit{Do the enforcement fines vary across the articles referenced in the enforcement decisions?}}
\item{\Qc:~\textit{How well the recent GDPR fines can be predicted in terms of
    basic available (i)~meta-data and (ii)~textual traits derived from the
    enforcement decisions?}}
\end{description}

It is difficult to make prior speculations about potential answers to the
questions. Regarding \Qa,~it can be expected that A5 is frequently referenced as
it specifies the overall lawfulness condition for processing personal data. But
beyond that, the GDPR contains as many as 99 articles, many of which may be used
to justify sanctions. As for \Qb,~it could be hypothesized that information
security lapses would yield particularly severe penalties; data breaches, in
particular, have often been seen as a major deterrent of the GDPR for
companies~\cite{Neto21}. With respect to \Qc,~there is a more practical
motivation: by knowing whether the penalties are predictable by machine learning
techniques, a starting point is available for providing further insights in
different practical scenarios. These scenarios include the automated archival of
enforcement decisions, information retrieval, designation of preventive
measures, and last but not least, litigation preparations.

From a data mining perspective, an answer to \Qc~further paves the way for
better understanding whether the manual labor required to construct meta-data
from unstructured administrative documents is necessary for predictive
tasks---or whether the documents are sufficient themselves. To this end, the
paper uses meta-data and text miming features extracted from the decision
documents. As such, only black-box predictions are sought; the goal is not to
make any legal interpretations whatsoever. The black-box approach also places
the paper into a specific branch of existing research dealing with legal and
administrative documents. After a brief further motivation for the regulation's
enforcement in Section~\ref{sec: background}, the related branch of work is
discussed in Section~\ref{sec: related work}. Thereafter, the paper's structure
is straightforward: the dataset and methods are elaborated in Sections~\ref{sec:
  data} and \ref{sec: methods}, results are presented in Section~\ref{sec:
  results}, limitations are discussed in Section~\ref{sec: limitations}, and
conclusions are summarized in the final Section~\ref{sec: conclusion}.

\section{Background}\label{sec: background}

There are many different viewpoints for approaching the enforcement
penalties. One possibility would be to focus on non-compliant products. As
noted, however, it is difficult to make generalizations due to the variety of
products processing personal data. Another viewpoint is to focus on the
regulators instead of the regulation; on the administration of the GDPR by
national DPAs and their EU-level coordination institutions. This viewpoint is
suitable for the present purposes. In contrast to domain-specific studies on the
GDPR and conformance with it, relatively little has also been written from this
administrative viewpoint.

\begin{figure}[t]
\centering
\includegraphics[width=\linewidth, height=10cm]{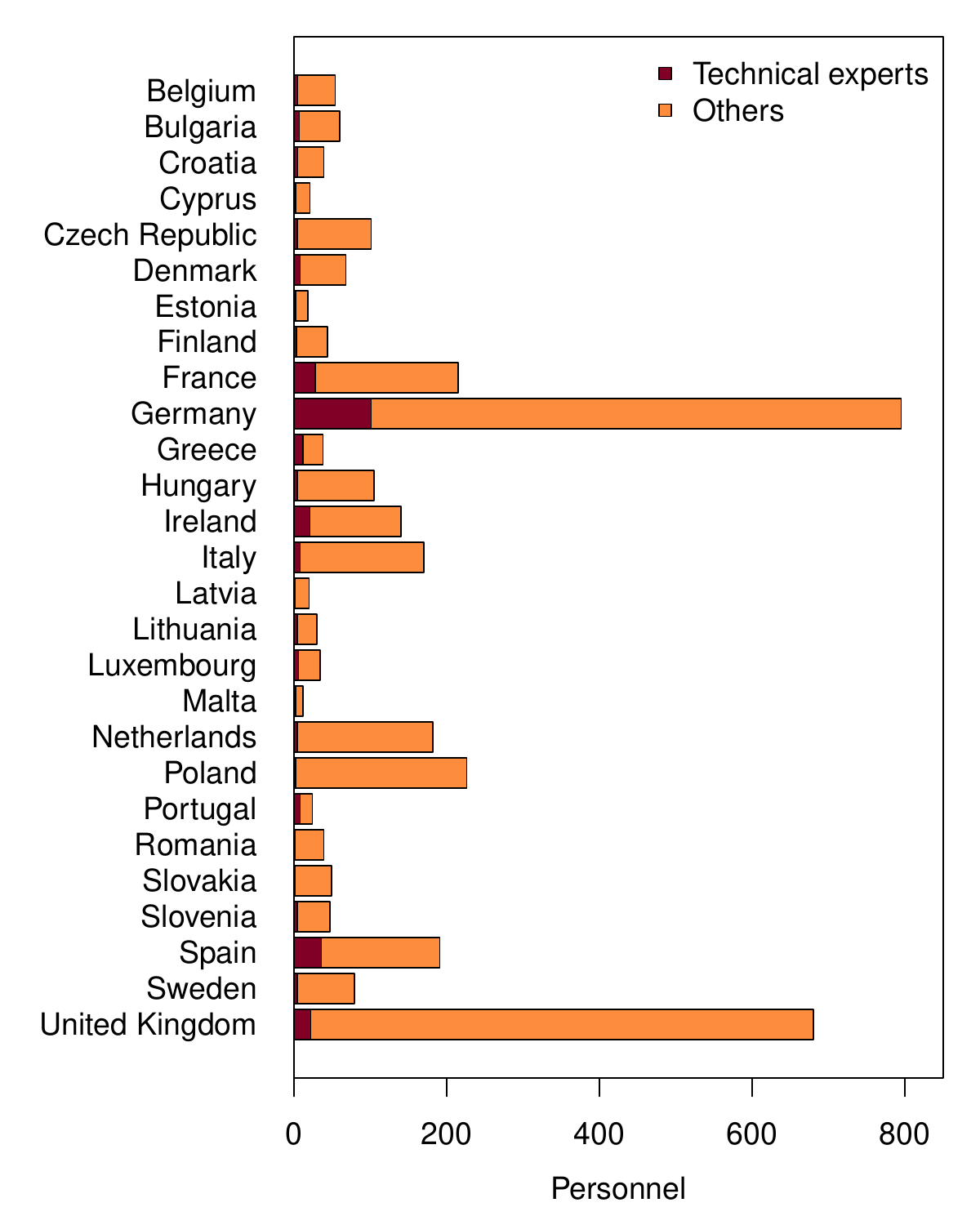}
\caption{Personnel Employed by DPAs in Selected European Countries in 2020
  (full-time employees addressing technical issues in private sector data
  processing; based on estimates reported in \cite{Brave20a})}
\label{fig: brave}
\end{figure}

In contrast to Directive 95/46/EC, Regulation (EU) 2016/679, the GDPR, is a
regulation; it is binding throughout the EU with only a minimal space for
national adaptations.\footnote{~Either the GDPR or comparable national laws have
been adopted also by countries participating in the EU's internal market via the
European Economic Area (EEA) treaty. For brevity, however, this detail is
omitted in what follows.} In practice, only a few articles in the GDPR provide
some but limited room for national maneuvering; these include A6 with respect to
relaxation in terms of other legal obligations or public interests, A9 in terms
of sensitive data, and A10 regarding criminal matters. Thus, in general, this
particular legislation should be interpreted and enforced uniformly through the
European Union by national data protection authorities whose \textit{formal}
powers are defined in A58. In practice, however, already the resources and thus
the \textit{actual} power for enforcement have varied across the member
states~\cite{Bennett18a, Custers18}. Although the budgets of the DPAs have
increased after the enactment of the GDPR in 2016 and its later enforcement in
2018, the resources remain scarce according to many critics. The resourcing
obstacles were also acknowledged by the European Commission in its 2020 review
of the GDPR's implementation. Accordingly, there is still a ``need for data
protection authorities to be equipped with the necessary human, technical and
financial resources to effectively carry out their tasks''~\cite{EC20a}. Besides
plain budgetary aspects, the lack of human resources is worth emphasizing. As
can be concluded from Fig.~\ref{fig: brave}, the amount of personnel employed by
national DPAs vary greatly across Europe. There is also an apparent lack of
engineers and other technical specialists employed by the national DPAs. Most of
the current employees are civil servants specialized to administration,
jurisprudence, and related non-technical areas of expertise.

By hypothesis, this evident cross-country variance reflects itself also in terms
of the enforcement fines imposed by the national DPAs. There are good reasons to
expect that the hypothesis is true. For instance, the enforcement of the GDPR
has been continuously criticized by some public authorities and pundits
alike. In addition to the lack of resources and the so-called ``one-stop-shop''
system, there are many other tenets in the criticism, including a lack of
transparency and cooperation between DPAs, diverging legal interpretations,
cultural conflicts, prioritization inconsistencies, old-fashioned information
systems, and general over-tolerance or even reluctance to enforce laws
~\cite{Casarosa20, ReneAusloos20, Ruohonen20COURT}. Although already the legacy
Directive 95/46/EC established the autonomy of DPAs, data protection issues have
also frequently prompted different bureaucratic conflicts and power struggles
within national public administration systems~\cite{Flaherty86,
  Yesilkagit11}. Fragmentation in terms of the national adaptations of the 1995
directive was also a well-recognized problem~\cite{Pearce98}. The interplay
between national and EU-level administration has caused additional problems for
European data protection~\cite{Lind14}. These are hardly unique issues in the
European Union in general.

Therefore, these problems and the cross-country incoherence should not be
overemphasized. Similar problems exist in many other policy areas in the
  EU, including closely related ones such as cyber
  security~\cite{Ruohonen20EJSR, Ruohonen16GIQ} and product
  safety~\cite{Ruohonen21SF} administration. Given that the GDPR contains
information security requirements (as specified particularly in A5 and A32),
data protection also aligns with cyber security in Europe. From this viewpoint,
the GDPR is best portrayed as a one piece in the EU's broader judicial framework
dealing with cyber security, trust, privacy, electronic commerce, and even cyber
crime~\cite{Mantelero21, Ventrella20, WickiBirchler20}. The same
applies to the enforcement and administration of the corresponding
laws. According to recent interviews of some key policy stakeholders, indeed,
the role played by DPAs is ranked high also with respect to cyber
security~\cite{SterliniMassacci20}. Given this broader viewpoint, perhaps more
than anything else, the GDPR's early enforcement problems reflect the general
administrative and political problems in the EU. And given these problems in
turn, it may be that comprehensive enforcement will be done in court rooms
through class actions~\cite{Casarosa20}. To this end, Directive 2019/2161 has
already been enacted for allowing collective redress for consumers and their
representatives.

\section{Related Work}\label{sec: related work}

Legal mining---for lack of a better term---has emerged in recent
years as a promising but at times highly contested interdisciplinary field that
uses machine learning techniques to analyze various aspects related to
law~\cite{Dyevre19, Leith16}. Although the concrete application domains vary,
case law and court cases are the prime examples already because these constitute
the traditional kernel of legal scholarship. Within this kernel, existing
machine learning applications range from the profiling of judges' personal
characteristics~\cite{Hausladen20, Wang20}, which may be illegal in some
European countries~\cite{CITIP20a}, to the prediction of decisions made by the
European Court of Human Rights~\cite{LiuChen17, Medvedeva19}, the Court of
Justice of the European Union~\cite{Moodley19}, and related chief judicial
authorities in Europe and elsewhere. These case law examples convey the two
traditional functions of applied machine learning; exploratory data mining and
forecasting.

Oftentimes, the legal mining domain is further motivated by a traditional
rationale for empirical social science research: to better understand trends and
patterns in lawmaking and law enforcement; to contrast these with legal
philosophies and theories; and so forth. Besides the goal of ensuring consistent
rulings~\cite{Wang20}, the rationale extends to public administration: machine
learning may ease the systematic archiving of legal documents and the finding of
relevant documents, and, therefore, it may also reduce administrative
costs~\cite{Chatwal17}. These administrative aspects reflect the goal of
building ``systems that assist in decision-making'', whereas the predictive
legal mining applications seek to build ``systems that make
decision''~\cite{Nissan18}. At the risk of a slight overgeneralization, it can
be said that the latter systems mostly equate to supervised machine learning
models, whereas the assisting systems usually operate with different,
law-specific information retrieval techniques. Particularly the information
retrieval techniques constitute the backbone in many legal experts systems in
practical use. While the present work belongs to the predictive domain, it
should be remarked that fully autonomous predictive systems are still rare in
law enforcement---and remain highly controversial.

Relying on distinct argumentation styles in legal reasoning~\cite{Atkinson20,
  Dyevre19}, the information retrieval systems extract and quantify textual data
from legal documents into structured collections with a predefined logic and
semantics~\cite{Bhuiyan19, Holzenberger20, Sleimi19}. To gain a hint about the
extraction, one might consider a legal document to contain some facts, rights,
obligations, and prohibitions, statements and modalities about these, and so
forth. This illustration helps to understand why a concept of legal
linguistics~\cite{Vogel18} is also sometimes used to describe the information
retrieval~approaches.

Although applications related jurisprudence are in the mainstream, it is worth
noting that similar techniques have also been used to extract requirements for
software and systems in order to comply with the laws from which a given
extraction is done~\cite{Sleimi19}. Driven by the genuine interest to facilitate
collaboration between lawyers and engineers in order to build law-compliant
software and systems~\cite{vanDijk18}, this rationale has been particularly
prevalent in the contexts of data protection and privacy. For instance, previous
work has been done to extract requirements from the Health Insurance Portability
and Accountability Act in the United States~\cite{Breaux06}. Against this
backdrop, it is no real surprise that data extraction has been applied also for
laws enacted in the EU. In particular, there are various existing works on
identifying requirements from the GDPR, including those based on manual
inspection and user stories~\cite{Bartolini19, Hjerppe19a}, ontologies and
information retrieval~\cite{Palmirani18, Tamburri20}, and formal
analysis~\cite{Arfelt19}. By and large, these works have concentrated on
providing a better understanding of the GDPR for technical implements and their
compliance.  Many---but not all~\cite{Meurisch21}---of these previous works also
limit themselves to requirements for technical implementations, omitting the
organizational requirements, such as the mandate to designate data protection
officers specified in A37.  As already noted, a different viewpoint is available
by focusing on the administration. Thus far, furthermore, the regulation's
enforcement has received only minimal attention. Apart from a few short
commentaries~\cite{Barrett20, Erickson19, ReneAusloos20}, no directly comparable
previous research seem to exist.

Finally, it should be emphasized that the decision documents released by the
national DPAs should not be strictly equated to law-like legal documents. On one
hand, the nature of these documents separates the present work from the
traditional applications in the legal mining domain; on the other hand, these
also enlarge the scope to which the work can be compared. For instance, highly
similar machine learning and information retrieval techniques have been used to
analyze privacy policies of software and systems~\cite{Harkous18,
  Lippi19}. Besides aligning closely with the GDPR's
requirements~\cite{Hjerppe20IWPE}, these also resemble the decision documents in
that neither a universal format nor well-defined semantics exist for
representing privacy policies. On that note, the dataset used should be
described in more detail.

\section{Materials}\label{sec: data}

\subsection{Dataset}\label{subsec: dataset}

The EU has not established a common database for archiving and cataloging the
GDPR enforcement decisions made by the DPAs. Although the European Data
Protection Board (EDPB), which supervises the national DPAs and coordinates
pan-European data protection activities, has recently established a specific
register for the ``one-stop-shop'' decisions made under A60~\cite{EDPB20a}, a
unified, comprehensive, and robust data source is lacking for the national
decisions made and the fines imposed by the~DPAs.

To patch this practical but important administrative limitation, several online
data collections have recently been established by non-governmental
organizations, companies, and others~\cite{NPLP20, noyb20a,
  PrivacyAffairs20a}. Also the dataset for the present work is based on an
online collection maintained by an international law firm for archiving many of
the known GDPR enforcement cases~\cite{enforcementtracker20}. Given that
annotation and labeling are often encountered problems for unstructured
collections of law-related documents~\cite{Lippi19, Sharafat19}, there is a
simple but important benefit from using the collection: each archived
enforcement case is accompanied by ready-made meta-data supplied by the firm as
well as a link to the corresponding decision from a data protection authority.

\subsection{Data Quality}\label{subsec: data quality}

The dataset is on the small side ($n = 294$), but still sufficient for
statistical inference and machine learning computations. Rather than the sample
size, the downsides of the dataset collected are elsewhere. In addition to
potentially missing cases due to a lack of publicly available information, the
archival material is unfortunately incomplete in many respects. The reason
originates from the incoherent reporting practices of the national data
protection authorities. Therefore, all available cases were obtained from the
online collection, but the following steps (see Fig.~\ref{fig: sample}) were
followed to construct a sample for the empirical analysis:

\begin{figure}[t]
\centering
\includegraphics[width=7.5cm, height=6cm]{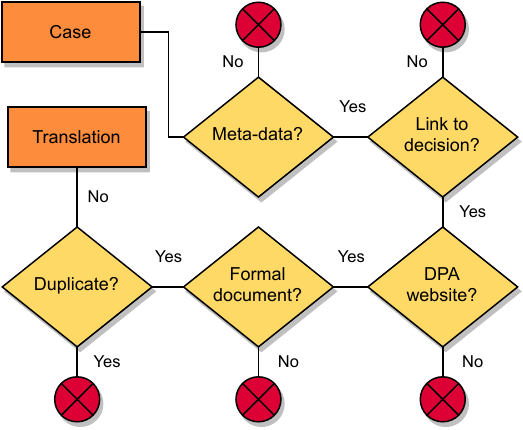}
\caption{Sample Construction}
\label{fig: sample}
\end{figure}

\begin{enumerate}
\itemsep 2pt
\item{To maintain coherence between the three research questions, only those
  cases were included that had both meta-data and links to the decisions
  available. In terms of the former, some cases lacked meta-data about the fines
  imposed, the particular articles referenced in the decisions, and even links
  to the decisions.}
\item{To increase the quality of the sample, only those cases were included that
  were accompanied with more or less formal documents supplied on the official
  websites of the European data protection authorities. By implication, those
  cases are excluded whose archival material is based online media articles,
  excerpts collected from annual reports released by the authorities, and
  related informal or incomplete sources.}
\item{If two or more cases were referenced with the same decision in the online
  archive, only one decision document was included but the associated meta-data
  was unified into a single case by merging the articles references and totaling
  the enforcement fines imposed.}
\item{Following recent research~\cite{Ruohonen20MISDOOM}, all national decisions
  written in languages other than English were translated to English with Google
  Translate. In general, such machine translation is necessary due to the
  EU-wide focus of the forthcoming empirical analysis.}
\end{enumerate}

\begin{figure}[th!b]
\centering
\includegraphics[width=7cm, height=7cm]{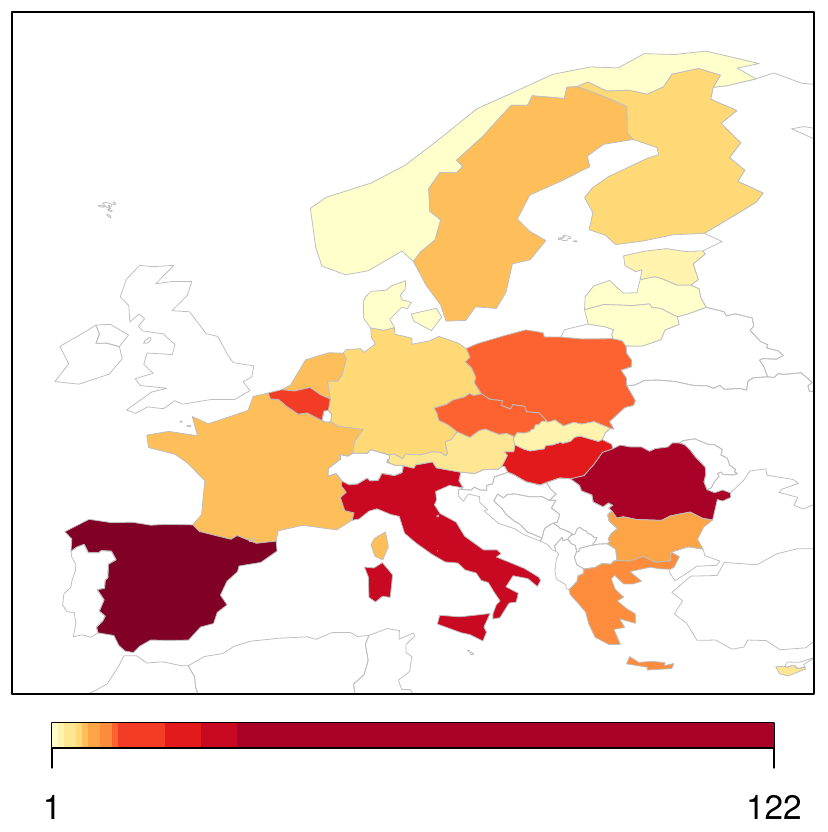}
\caption{Countries of Origin}
\label{fig: countries}
\end{figure}

Given these restrictions, the $n = 294$ cases in the sample amount to about 73\%
of all cases archived to the proprietary online archive at the time of the data
collection (24 September, 2020). The coverage is thus good even with the
exclusions. However, it should be noted that the quality of the sample is not
optimal. Two points warrant a brief discussion in this regard. First, partially
due to the data availability issues, the sample is not spatially balanced across
Europe. As can be observed from Fig.~\ref{fig: countries}, many of the
enforcement fines in the sample were made in Spain, whereas the decisions of
German data protection authorities are likely underrepresented in the
sample. Germany is also otherwise an exception since there are multiple German
DPAs operating at the state level instead of a single data protection authority
at the national level.

\begin{figure}[th!b]
\centering
\includegraphics[width=\linewidth, height=4cm]{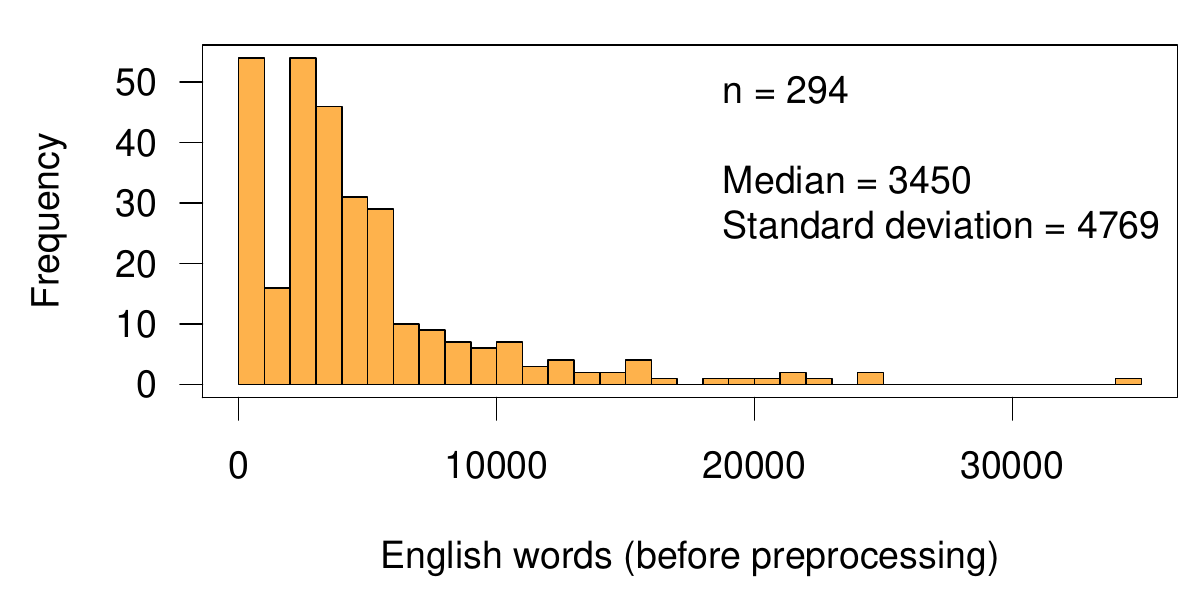}
\caption{Decision Document Lengths}
\label{fig: lengths}
\end{figure}

Second, the authorities in some countries have released highly detailed and
rigorous documents about their decisions, while some other authorities have
opted for short press releases. Although the length of a document does not
necessarily reveal its quality, in the present context, the large variance seen
in Fig.~\ref{fig: lengths} illustrates the lack of rigor present in some
documents; when imposing fines, which may be substantial under the GDPR, a
decision justified with a few thousand words does not seem optimal. It is
also worth remarking that most of the documents were supplied in the portable
document format (PDF) and informally signed by the authorities (of all
  documents retrieved, about 77.9\% were PDF files; the rest are plain texts
  appearing on the DPAs' websites). However, scanned PDF documents had to be
excluded due to the automatic data processing. For instance, the scanned PDF
documents used in Portugal were omitted (cf.~Fig.~\ref{fig: countries}). These
data quality issues and their implications are further discussed in
Section~\ref{sec: limitations}. For the time being, it suffices to again
stress that the quality issues are related to the general administrative and
political shortcomings in the EU.

\subsection{Preprocessing}\label{subsec: preprocessing}

The textual aspects for \Qc~are derived from the translated decisions. A
conventional ``bag-of-words'' approach is used for extracting the features. This
choice is justifiable due to the nature of the dataset. The machine-translation,
which is necessary for a EU-wide analysis, largely prevents robust use of
semantic approaches based on word embeddings, part-of-speech (PoS) tagging, and
related techniques. Furthermore, the decision documents vary greatly across
Europe in terms of style, conventions, format, and other linguistic elements. In
essence, each European DPA tends to use a distinct style and convention for
documenting its decisions. This variance further implies that the information
retrieval techniques developed in the legal mining domain cannot be readily
applied.

Nevertheless, some preprocessing is still necessary. Nine steps were used for
the task. To begin with, (1) all translated decision documents were lower-cased
and (2) tokenized according to white space and punctuation characters; (3) only
alphabetical tokens recognized as English words were included; (4)~common and
custom stopwords were excluded; (5) tokens with lengths less than three
characters or more than twenty characters were excluded; and (6) all tokens were
lemmatized into their common English dictionary forms. A common natural language
processing library~\cite{NLTK20} was used for this processing together with a
common English dictionary~\cite{hunspell20}. In addition to the common stopwords
supplied in the library, the twelve most frequent tokens were used as custom
excluded stopwords: \textit{data}, \textit{article}, \textit{personal},
\textit{protection}, \textit{processing}, \textit{company}, \textit{authority},
\textit{regulation}, \textit{information}, \textit{case}, \textit{art}, and
\textit{page}.

After these initial steps, (7)~five separate corpora were constructed by using
$k$-grams with $n = 1, \ldots, 5$. These contain sequences of adjacent
lemmatized tokens; for instance, the phrase \textit{condicio sine qua non}
yields three $2$-grams: \textit{condicio sine}, \textit{sine qua}, and
\textit{qua non}. In general, $k$-gram models are commonly used in text
mining as these often improve predictions and ease interpretation. The legal
mining domain is not an exception in this regard~\cite{Hausladen20,
  Medvedeva19}. After the construction of these five corpora, (8)~each one was
pruned by excluding those $k$-grams that occurred in a given corpus
only once. Finally, (9)~term frequency inverse document frequency (TF-IDF)
scores were calculated for the $k$-grams in each corpus (for the exact
formula used see~\cite{Ruohonen18TIR}). In general, TF-IDF is often preferred as
it penalizes frequently occurring terms. It is also worth remarking that other
common weighting schemes~(see, e.g., \cite{Fang04, JinChai05}) did not notably
change the empirical predictions reported.

\section{Methods}\label{sec: methods}

Descriptive statistics are used to answer to \Qa, and ordinary least squares
(OLS) to \Qb. Regarding the latter question, two OLS models are estimated: a
restricted one in which only the articles referenced in the decisions are
present, and an unrestricted one that includes rest of the meta-data. A
logarithm of the enforcement fines is used as the dependent variable in both
OLS regression models.

The restricted regression model equates to the conventional analysis-of-variance
(ANOVA). For the unrestricted OLS model, the additional meta-data aspects
include dummy variables for the following features: (i)~the \textit{year} of a
given enforcement case; (ii) the \textit{country} in which the given fine was
imposed; and (iii) the \textit{sector} of the violating organization. The last
feature was constructed manually by using five categories: individuals, public
sector (including associations, political parties, universities, etc.),
telecommunications, private sector (excluding telecommunications), and unknown
sector due to a lack of meta-data supplied in the online archive. Together
with an intercept, the unrestricted model contains $55$ independent variables.

\begin{figure}[t]
\centering
\includegraphics[width=7.5cm, height=14cm]{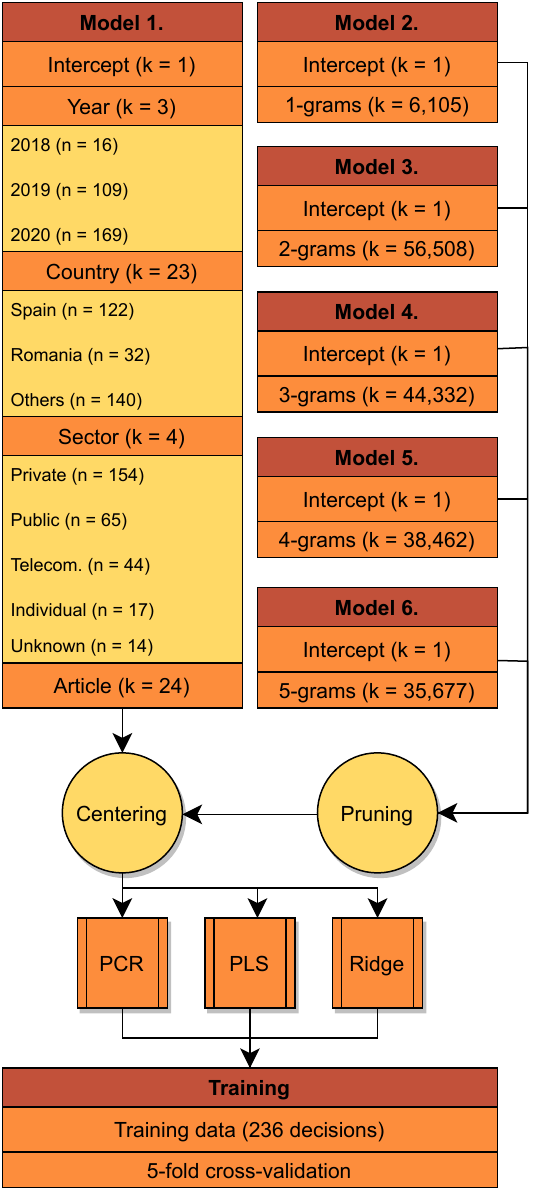}
\caption{Model Construction}
\label{fig: model}
\end{figure}

The question \Qc~requires a different strategy. The reason is sparsity: there
are only $294$ enforcement decisions, while each $k$-gram corpus contains
thousands of $k$-grams. In fact: even after the eight preprocessing steps noted
in Section~\ref{subsec: preprocessing}, there are over $35$ thousand features in
each $k > 1$ corpus (see Fig.~\ref{fig: model}). Fortunately, the problem is not
uncommon, and dimension reduction is the generic solution for addressing it. To
this end, each corpus was further pruned with the \texttt{nearZeroVar} function
available from the package~\cite{caret20} used for computation. It drops those
features that have only one unique value, as well as those features that have a
very few unique values whose frequency is large with respect to the second most
common value.

Then, three common dimension reduction methods for regression analysis are used:
principal component regression (PCR), partial least squares (PLS), and ridge
regression. In essence, PCR uses uncorrelated linear combinations as the
independent variables; PLS is otherwise similar but also the dependent variable
is used for constructing the combinations. Ridge regression is based on a
different principle: the dimensionality is reduced by shrinking some of the
regression coefficients toward zero. All three are classical and well-documented
regression methods (for summaries of the statistical background see
\cite{Hastie11} and \cite{Kiers07}). All are also widely used in applied
research \cite{Hemmateenejad07}. In general, all three methods are further known
to yield relatively similar results in applied work. Given these points, it is
more relevant to proceed by elaborating the practical computation than to
describe the methods themselves.

Thus, in terms of practical computation, the number of components for the PCR
and PLS models, and the shrinkage parameter for the ridge regression, is
optimized during the training while the results are reported with respect to a
randomly selected test set containing 20\% of the enforcement cases. Centering
(but not scaling) is used prior to the training with a $5$-fold
cross-validation. Computation is carried out with the \textit{caret}
package~\cite{caret20} in conjunction with the \textit{pls}~\cite{pls07} and
\textit{foba} \cite{foba08} packages. Although root-mean-square errors (RMSEs)
are used for optimization, the results are summarized with mean absolute errors
(MAEs) due to their straightforward interpretability. These are defined as the
arithmetic means of the absolute differences between the observed and predicted
fines in the test~set.

As for answering to \Qc~in general, each of the three regression estimators is
used to estimate six models (see~Fig.~\ref{fig: model}). The first contains the
meta-data features; the second and third models the pruned \text{$2$-gram} and
\text{$3$-gram} features; and so on. If the meta-data model outperforms the
textual feature models, at least one of the estimators should show smaller MAEs
compared to the MAEs from any of the fifteen models using the $k$-gram features.

\section{Results}\label{sec: results}

\subsection{Fines}

\begin{figure}[th!b]
\centering
\includegraphics[width=\linewidth, height=4cm]{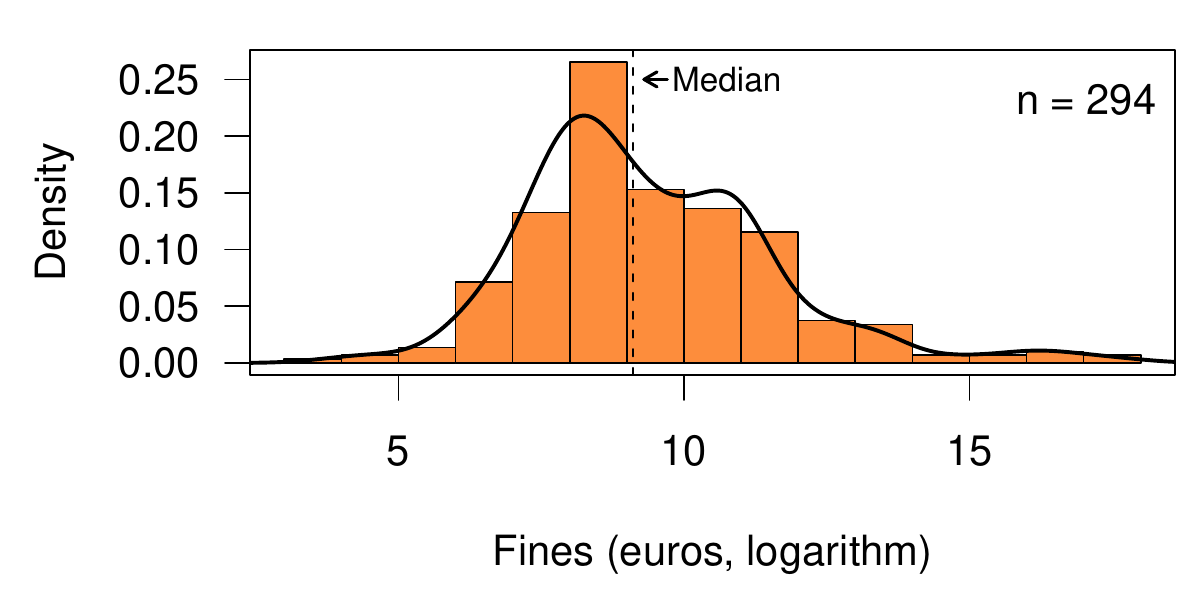}
\caption{Enforcement Fines in the Sample}
\label{fig: fines}
\end{figure}

The GDPR enforcement fines imposed vary greatly in the dataset. As can be seen
from Fig.~\ref{fig: fines}, a range from about $e^6$ euros to $e^{12}$ euros
capture the majority of the enforcement fines observed. This range amounts
roughly from about four hundred to $163$ thousand euros.

That said, the distribution has a fairly long tail; also a few large,
multi-million euro fines are present in the sample. Therefore, the sample cannot
be considered biased even though the restrictions discussed in
Section~\ref{subsec: data quality} exclude some of the largest enforcement
cases, including the announcements about the intention to fine the British
Airways and Marriott International by the Information Commissioner's Office in
the United Kingdom. Although these two excluded cases are (at least at the time
of writing) preliminary announcements, they are still illuminating in the sense
that both were about large-scale data breaches of consumer data. Given that data
breaches have been estimated to cause hundreds of millions (or more) of economic
and societal losses~\cite{Edwards16, Poyraz20}, even the few large fines in the
sample are clearly on the small side. This point reinforces the earlier remarks
about the enforcement problems.

\subsection{Articles}

\begin{figure}[th!b]
\centering
\includegraphics[width=\linewidth, height=10cm]{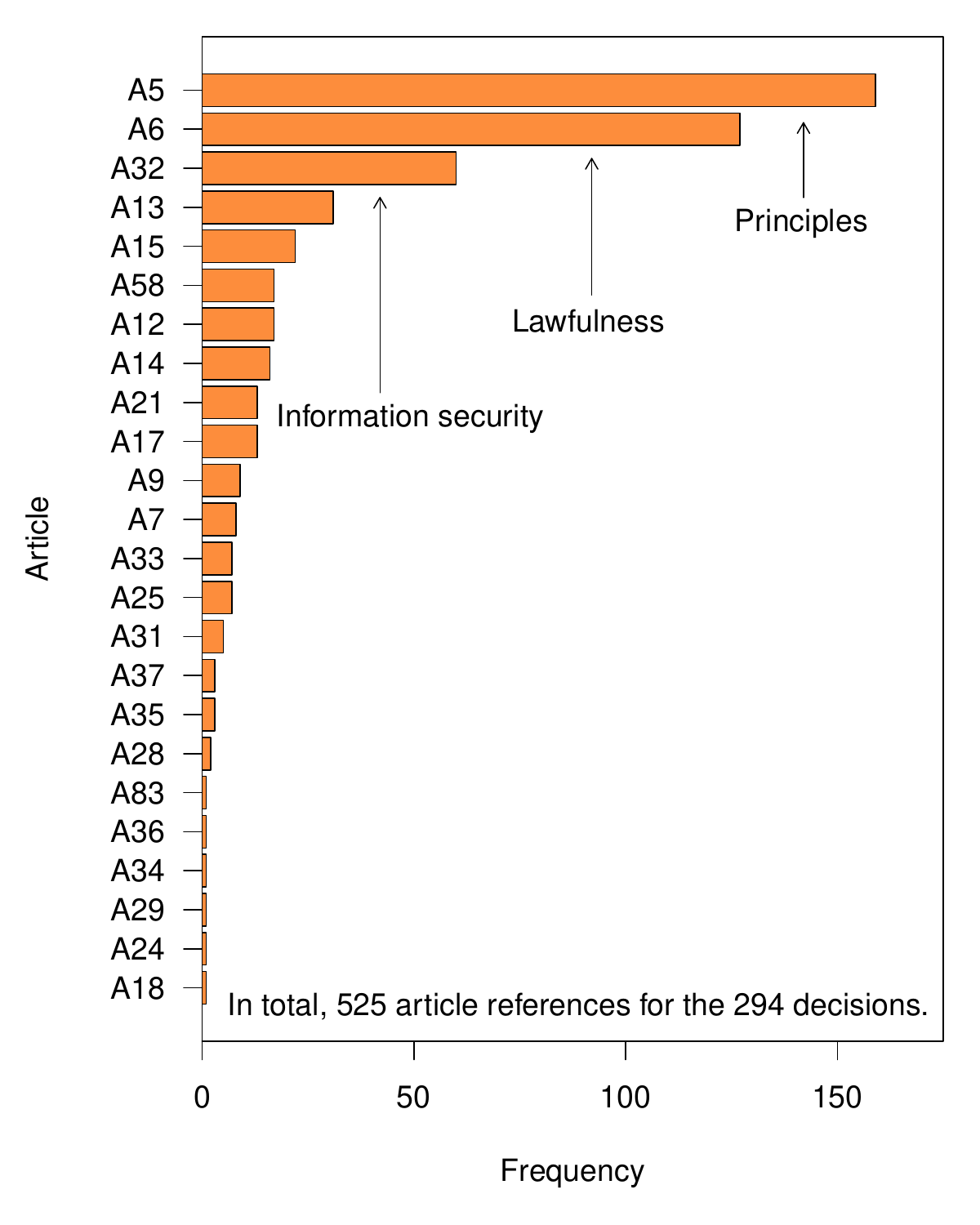}
\caption{Articles Referenced in the Enforcement Cases}
\label{fig: articles}
\end{figure}

Articles A5 and A6 have been the most frequently referenced ones in the
enforcement decisions (see~Fig.~\ref{fig: articles}). This observation is not
surprising; these two articles are perhaps the most fundamental ones among the
ninety-nine articles laid down in the GDPR. Article A5 specify the
accountability criterion and the mandate to be able to demonstrate
compliance. These are fundamental practically for all software products
processing with personal data~\cite{Hjerppe19a}. Article A6, in turn, specifies
the six conditions under which the lawfulness of processing personal data can be
established in the EU under the GDPR. Thus, it is no real wonder that as many as
67\% of the enforcement decisions have referenced either A5, A6, or both of
these.

Article~A32, which addresses the security of processing personal data
explicitly, has been the third most frequently referenced article in the
decision documents. Given that particularly the recital (f) in A5 align with
A32~\cite{Ventrella20}, it can be concluded that many of the decisions have
dealt with data breaches and other security lapses. When taking a look at the
twenty-five $2$-grams with the highest TF-IDF scores, different security issues
are indeed apparent; \textit{black list}, \textit{technical organizational} and
\textit{appropriate technical} (which both refer to A5), \textit{security
  breach}, \textit{unauthorized access}, \textit{security measure},
\textit{security policy}, and so forth.

\begin{figure*}[th!b]
\centering
\includegraphics[width=\linewidth, height=10cm]{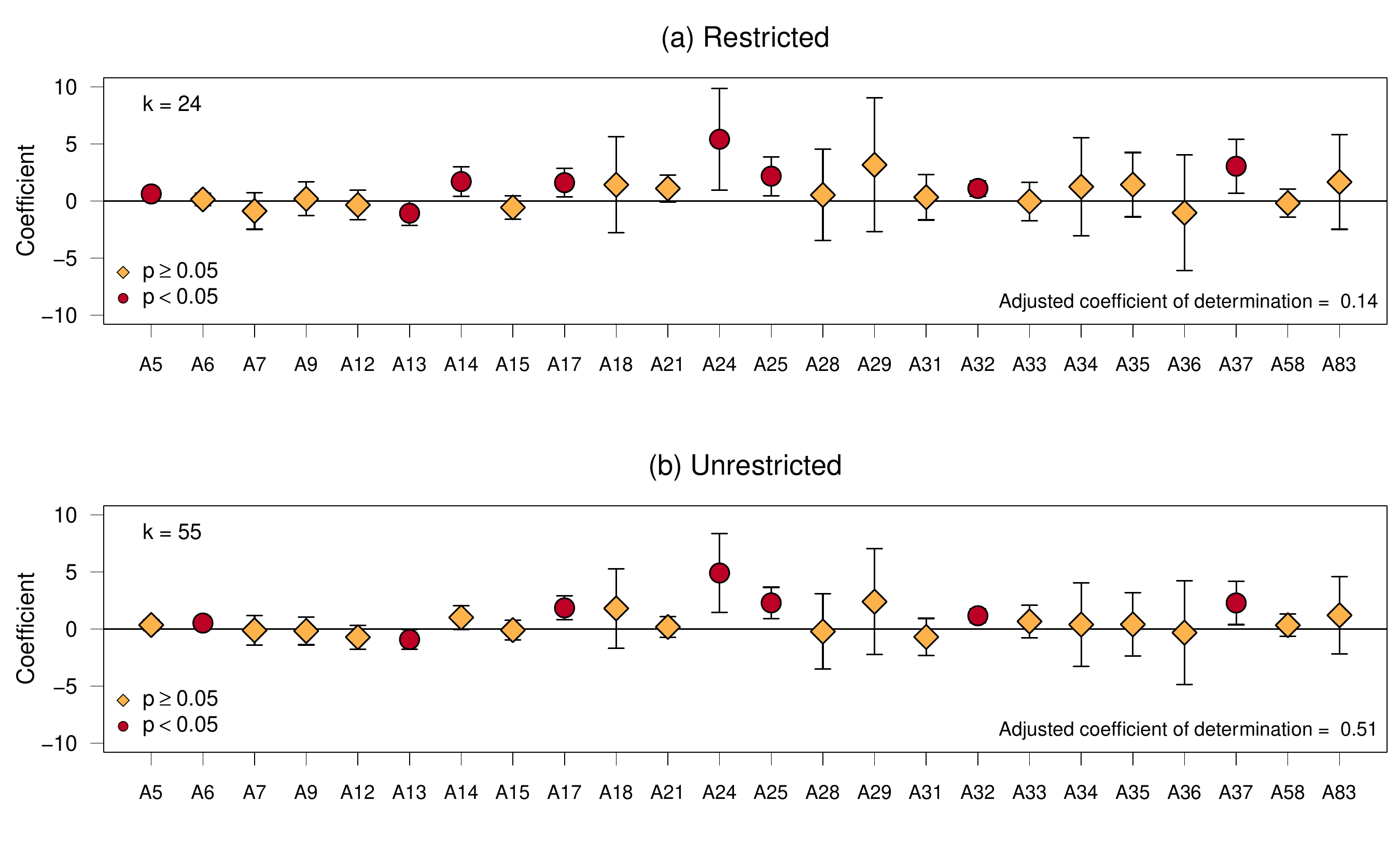}
\caption{Enforcement Fines Across Articles (logarithm, OLS, 95\% CIs)}
\label{fig: ols}
\end{figure*}

In addition, many references have also been made to numerous other articles in
the GDPR. As many as $31$ references have been made to A13 and $22$ references
to A15. The former specifies the informing obligations to data subjects, whereas
the latter defines the conditions under which they can access their personal
data. In general, these references reflect the criticism about the
non-compliance of many organizations with respect to their respect of the new
rights granted to individuals~\cite{Mahieu18}. Furthermore, more than seven
references have been made to A58 (the powers granted to DPAs), A12 and A14
(transparency requirements), A21 (the right to object), A17 (the right to
erasure), A9 (sensitive personal data), and A7 (the conditions for
consent). Particularly the seventeen references to A58 are worth emphasizing; it
seems that some organizations have been also unwilling to cooperate with public
authorities. In this regard, it is further worth remarking the references made
to the obligations to designate data protection officers (A37), conduct impact
assessments (A35), and consult supervisory authorities (A36), to name three
examples. Furthermore, less frequent references have been made in the decisions
to numerous other articles. Interestingly, though, no references have been made
to A22 (the right to object automatic decision-making that have legal
consequences for data subjects). This observation again pinpoints toward the
diverging legal interpretations in Europe~\cite{Suksi20}. But all in all, as a
whole, the GDPR articles referenced hint that the regulation's full scope is
slowly being enforced by the data protection authorities. Indirectly,
  the references to articles such as A35, A36, and A37 further hint that DPAs
  are also using their soft power for improving data protection. In addition to
  the enforcement decisions as a deterrent against poor practices, such soft
  power includes public relations, promotion of instructions and guidelines,
  raising of awareness, and other things.

Turning to the regression analysis, the OSL estimates are summarized in
Fig.~\ref{fig: ols}. There are three points worth making about the
estimates. First, the regression coefficients are highly similar between the
restricted model including only the articles and the unrestricted model
containing all available meta-data. In addition to the similarity in terms of
magnitude, only three coefficients differ between the two models with respect to
statistical significance at the conventional level. Second, the overall
performance is even surprisingly good for the unrestricted model; the adjusted
$R^2$ is as high as $0.51$. Given that the unrestricted model yields a value of
$0.14$, much of the performance is attributable to the other three meta-data
features. Of these features, none of the dummy variables are statistically
significant for the sector of an infringing party. Hence, the year of
enforcement and the country of origin are particularly relevant for explaining
the overall variation in the enforcement fines. This supports the earlier
discussion about cross-country variation in the GDPR's enforcement and the
administration of data protection in general. Third, none of the coefficients
forcefully stand out in terms of their magnitudes. When looking at the
coefficients with relatively tight confidence intervals (CIs), it is evident
that variation is present but the magnitude of this variation is not
substantial. Most of the coefficients remain in the range $[-5, 5]$. It is
particularly noteworthy that in both models the coefficients for A32 (the
information security requirements) are statistically significant and have a
positive sign, but with only modest magnitudes. The small magnitudes apply also
to A5. The observation is generally surprising, given that data breaches could
be expected to yield particularly severe penalties. But according to the
dataset, this expectation does not hold ground.

\subsection{Predictions}

\begin{figure}[th!b]
\centering
\includegraphics[width=\linewidth, height=10.5cm]{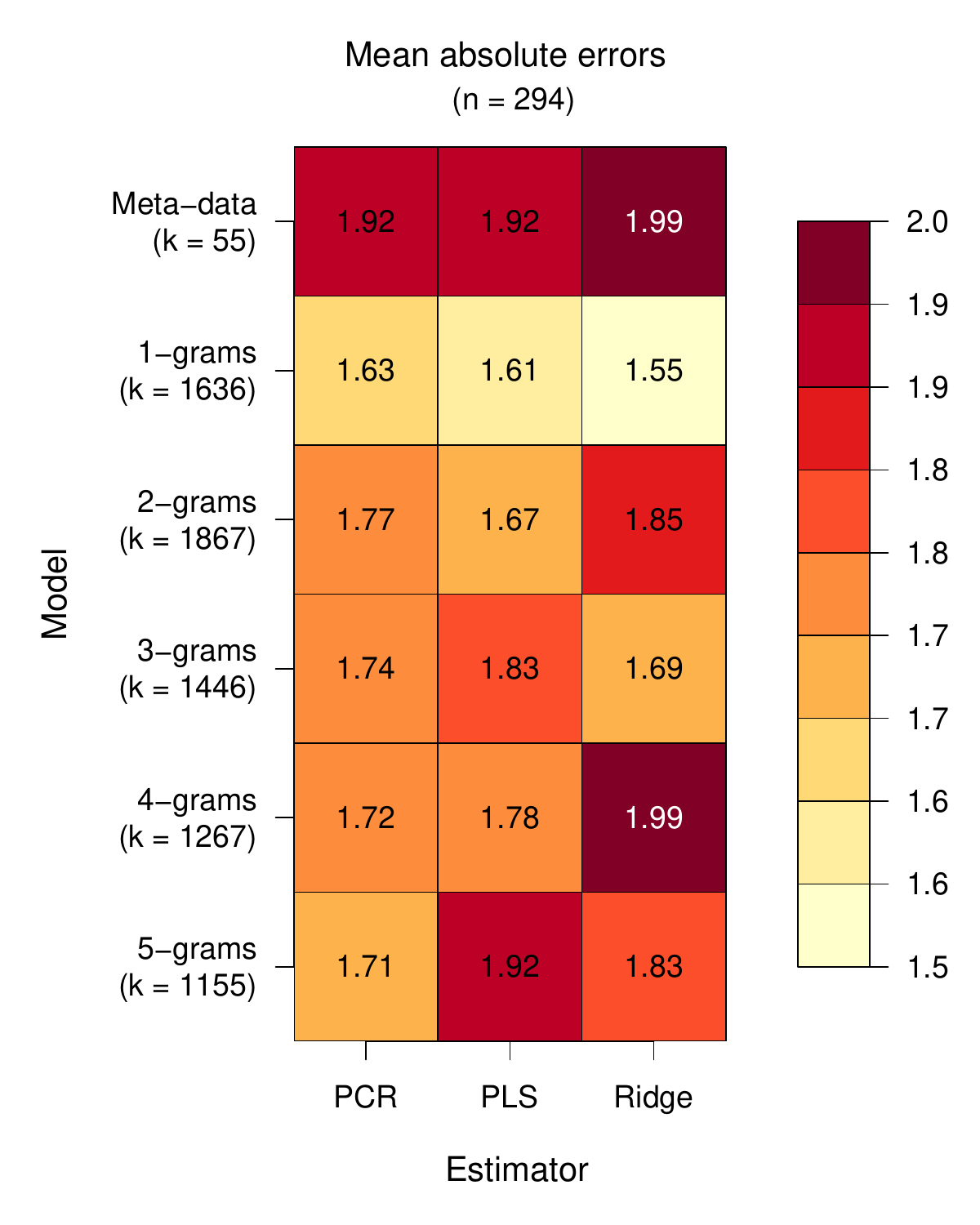}
\caption{Prediction Performance (MAEs)}
\label{fig: maes}
\end{figure}

\begin{figure}[th!b]
\centering
\includegraphics[width=\linewidth, height=5cm]{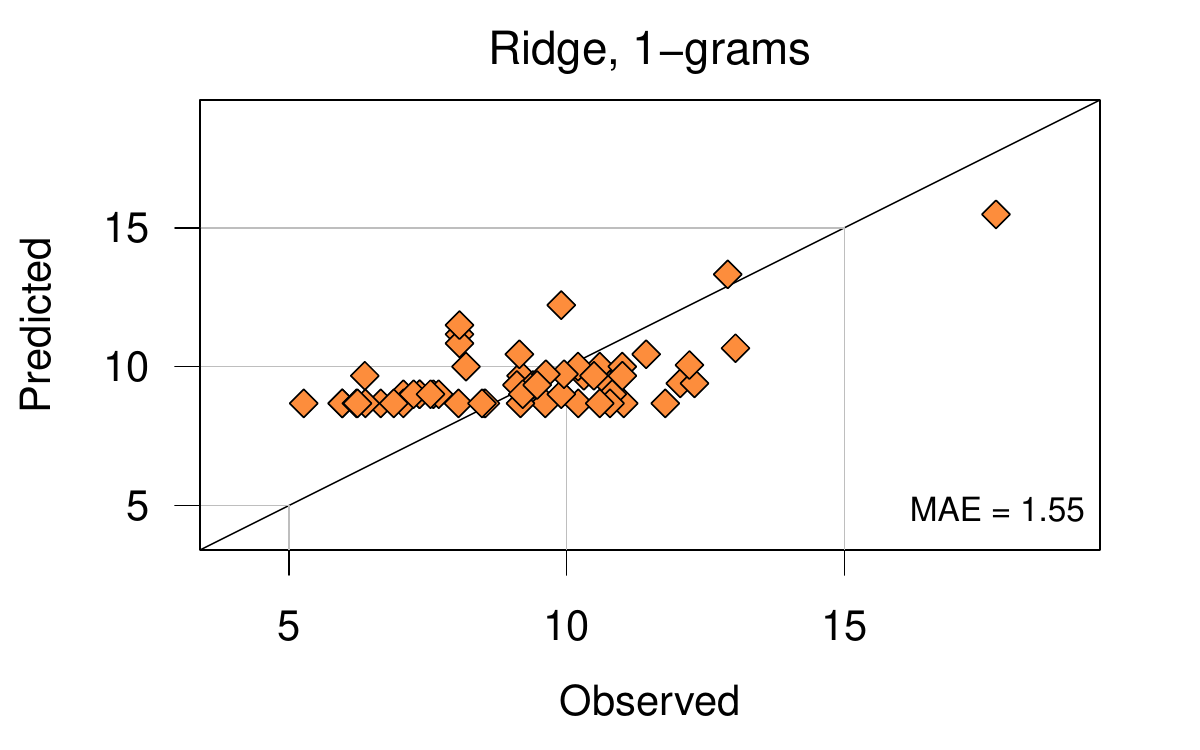}
\caption{Observed and Predicted Values in the Test Set}
\label{fig: pred}
\end{figure}

The results from the cross-validated predictions are summarized in
Fig.~\ref{fig: maes}. It shows the mean absolute errors across the $48$ models
trained. These errors are small. Given that all MAEs are below $e^2$ euros, the
predictions are generally decent enough. Another point worth remarking is that
Ridge regression with $1$-grams outperforms all other models. Therefore, the
answer to \Qc~is twofold: while meta-data gives decent predictions, the
mechanical black-box textual features yield slightly better ones. The estimates
seem acceptable also upon a close visual examination. For instance, in
Fig.~\ref{fig: pred}, even the outlying large fine is estimated with a
reasonable error. Adding a dummy variable for it and re-estimating the
  models indicates only small improvements in the MAEs. But as will be soon
  noted in Section~\ref{sec: limitations}, outliers still remain a potential
  concern for the prediction of future enforcement fines.

\begin{figure}[th!b]
\centering
\includegraphics[width=\linewidth, height=10.5cm]{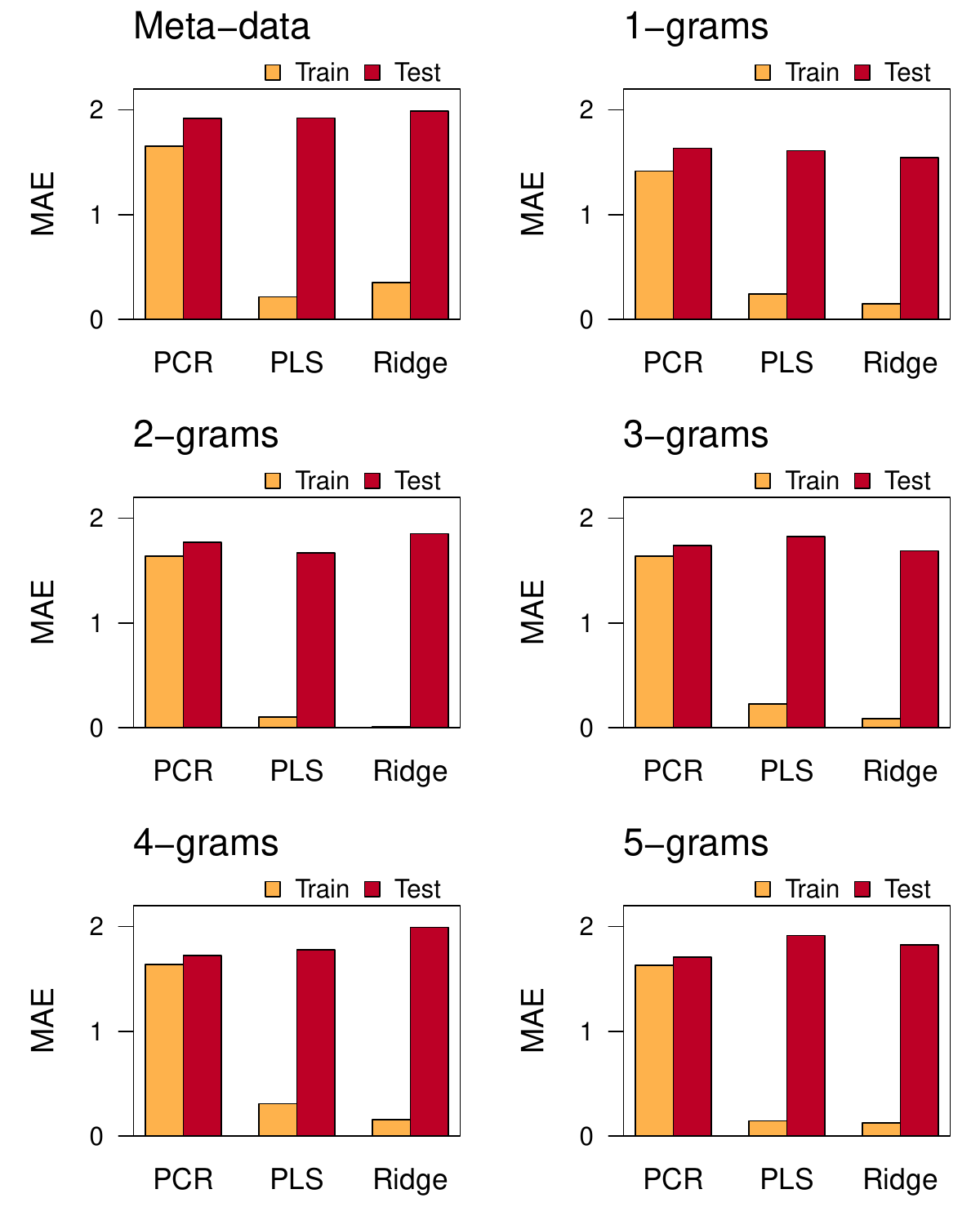}
\caption{Training and Testing (MAEs)}
\label{fig: train}
\end{figure}

The same concern can be raised also from the illustration in Fig.~\ref{fig:
  train}, which indicates potential problems in the training process, including
the possibility of over-fitting (the MAEs for the training refer to the best
cross-validated models). In other words, there are fairly large gaps in the
performance between the training and test sets. Though, these gaps apply only to
the PLS and Ridge regression estimators. Given that $1$-grams yield the best
performance also with the PCR estimator (see Fig.~\ref{fig: maes}), which does
not exhibit notable train-test gaps, the overall conclusion regarding \Qc~is not
threatened---for this particular dataset.

\section{Limitations}\label{sec: limitations}

Some limitations should be acknowledged. To raise the generality, the discussion
that follows addresses these together with points about automatic
decision-making systems for judiciary. Given the discussion in Section~\ref{sec:
  related work}, the paper aligns with the idea of systems making decisions; the
results presented can therefore be seen as an output from a prototype-like
automatic decision-making system.

The limitations can be further framed with the difficult concepts of
reproducibility and replicability (or repeatability). These concepts are often
used interchangeably. Sometimes, these are even defined in conflicting ways (see
for instance \cite{Repar20} versus \cite{Ruohonen15COSE}) even though the
intention remains the same. For the present purposes, repeatability can be
defined as a ``property of an experiment: the ability to repeat---or not---the
experiment described in a study'', whereas reproducibility is understood as ``a
property of the outcomes of an experiment: arriving---or not---at the same
conclusions, findings, or values'' \cite{Cohen18}. With some caveats, basic
repeatability should be possible by carefully following the steps described in
Sections~\ref{sec: data} and \ref{sec: methods}. But the text mining context
almost necessarily adds some caveats.\footnote{~In order to facilitate potential
  replications, the dataset used for the statistical computation is available
  online~\cite{Ruohonen21dataset}.}

Among the caveats is preprocessing. Even with the guidelines, achieving
\textit{perfect} repeatability may be difficult. But this difficulty comes from
a necessity: there are tens of thousands of $k$-grams in the corpora used to
answer to RQ$_3$, and even after preprocessing, the amounts are large and
require dimension reduction methods for regression analysis. This comes at a
cost for repeatability. A~related caveat is the machine-translation of the
decision documents to English. As a proprietary translation engine was used, it
is impossible to guarantee that exactly the same results would be obtained in
the future. With regard to the translations themselves, there is existing
discussion about the use of Google Translate in scientific and scholarly
applications~\cite{Daniele19, Groves15}. For the present purposes, it suffices
to briefly continue the discussion by noting that even small translation
mistakes may have exceptionally dire consequences in
judiciary~\cite{Scott21}. However: as only conventional TF-IDF weights
  were used, mistakes in translation semantics are a lesser concern for the
  present paper. That said, furthermore, a proprietary online database was used
to obtain the meta-data as well as to collect the decisions from the primary
sources. Even if a repeatable code would be supplied for the data retrieval, it
is impossible to guarantee that exactly same data would be retrieved due to the
third-party source.

But what do these repeatability concerns imply for ADM systems used in
judiciary? Clearly, in this context, nearly perfect repeatability should be
guaranteed as otherwise it becomes difficult, if not impossible, to challenge
and justify a system's decision. In this regard, there is an interesting recent
discussion about preprocessing and dimension reduction methods in automatic
decision-making systems used in judiciary: as preprocessing is absurd in the
legal mining domain because it may change cases extra-judicially, regularization
and related methods should be used instead~\cite{Bibal20, Boswell20}. Further
problems arise from the proprietary, closed source nature of most ADMs. In
general, it is notoriously difficult to audit such systems~\cite{Waltl18}. And
once again, the problem is not merely about auditing; it is about the use of
private sector systems for public sector services~\cite{Kerikmae20,
  Kuziemskia20, Suksi20}. Thus, the repeatability concerns are graver on the
side of ADMs due to the consequences to individuals subjected to the decisions
made by the systems.

Analogous concerns apply to the reproducibility of values, such as the
distribution of the fines in Fig.~\ref{fig: fines}. Here, the biggest concern is
generalizability. Although the about 73\% of decisions from the online archive
could be reasonably assumed to generalize toward all decisions in this
particular archive, these may not generalize toward the whole population of GDPR
enforcement decisions during the period studied. Again, the problem is
unavoidable because neither the national DPAs together nor the EU institutions
have provided a rigorous archive for all decisions made in Europe. In short:
because the statistical population remains unknown, generalizability cannot be
guaranteed.

These concerns about reproducibility of values translate into potential issues
in the reproducibility of findings, such as those in Fig.~\ref{fig: articles}
(RQ$_1$) and the ANOVA results in Fig.~\ref{fig: ols}~(RQ$_2$). The regression
predictions (RQ$_3$) are further threatened by other issues. For instance, the
dataset is not balanced across Europe~(see~Fig.~\ref{fig: countries}). There are
potential issues also with the sectoral breakdown (see Section~\ref{sec:
  methods}) because some countries (such as Finland) have excluded public sector
from the scope of A83. But all things considered, do these problems threaten the
reproducibility of conclusions, the answers to the three research questions?

By argument, the answer is negative: regardless of the repeatability and
reproducibility threats, a future study should find frequent references to A5
and A6 in particular~(RQ$_1$), variance of the enforcement fines across the
articles~(RQ$_2$), and decent predictions by conventional regression
methods~(RQ$_3$). Of these conclusions, the one given to RQ$_3$ is the most
contestable. The ADM context illustrates the issue better than the analysis
presented.

In general, ADMs used in judiciary have severe problems in reorienting
themselves according to changes in law and court practice~\cite{Suksi20}. Thus:
if future enforcement pushes the magnitudes toward billion-euro fines, say, the
predictions would be inaccurate at best and haphazard at worst. Yet the real
issue is not about potential inaccuracies. Predicting the GDPR enforcement fines
is as a sensible research question as any for scholarly work, but, throughout
this paper, an implicit question has lingered along: should automated systems
for determining fines be deployed in a society? If the answer is no, or even
maybe, there should be a thorough political discussion in the society.

\section{Conclusion}\label{sec: conclusion}

The following points summarize the answers reached:
\begin{enumerate}
\item{Based on a dataset constructed via a third-party collection---which is
  necessary because the public administrations involved have not been able or
  willing to provide adequate public data, thus casting the accountability and
  transparency of the enforcement into a somewhat dismal light---the articles
  related to the general principles (A5), lawfulness (A6), and information
  security (A32) have been the most frequently referenced ones in the recent
  enforcement decisions done by the public administrations (\Qa). The
  observation is not surprising. Article A5 is a go-to article due to its
  explicit responsibility dictate, and each one who computes with personal data
  must satisfy one of the legal basis in Article A6 (with few exceptions).}
\item{However, the enforcement fines are not forcefully larger or smaller for
  the decisions referencing A5, A6, and A32. Particularly A32 is surprising in
  this regard considering the harms caused by data breaches. But, in general,
  the enforcement fines do vary across the articles referenced (\Qb). That said,
  a slightly stronger statistical explanation is available by knowing the year
  of enforcement and the country in which a public authority making a decision
  is located. This statistical result reinforces the discussion in
    Section~\ref{sec: background} on the administrative problems in enforcing
    the GDPR.}
\item{Both the meta-data available from the third-party and the textual features
  extracted from the enforcement documents released by the responsible public
  authorities provide enough material for decent predictions of the enforcement
  fines (\Qc). The textual features seem to outperform the meta-data, suggesting
  the plausibility of using a black-box predictive system for foresight. The
  average error is less than ten euros.}
\end{enumerate}

There are a couple of prolific paths for further research. Besides potential
reproduction of the conclusions, first, it seems reasonable to argue that future
research should focus on providing a more nuanced analysis of the enforcement
decisions. A better understanding on the logic and arguments used in the
decision documents is necessary for moving forward with the domain of legal
linguistics described in Section~\ref{sec: related work}. Eventually, it may be
possible also in the GDPR context to build systems that assist in
decision-making---even if actual fine-imposing ADMs are seen as unachievable or
undesirable. The second path follows. By a fine-grained analysis, it should be
also possible to establish implicit compliance frameworks for
implementations. Fundamentally---like with all laws, the GDPR's enforcement
should not depend on sanctions but on acquiescence.

\section*{Acknowledgements}

This research was funded by the Strategic Research Council at the Academy of Finland (grant no.~327391).

\balance
\bibliographystyle{apalike}

\end{document}